\documentclass[aps,prd,noshowpacs,preprintnumbers,nofootinbib,floatfix,onecolumn]{revtex4}

\usepackage{bm}
\usepackage{latexsym}
\usepackage{dcolumn}
\usepackage{amsfonts,amssymb}
\usepackage{graphicx,epsfig}
\usepackage{psfrag}

\begin{document}


\title{Thermodynamics of  horizons from a dual quantum system}
\author{Sudipta Sarkar}
\email{sudipta@iucaa.ernet.in}
\author{T. Padmanabhan}
\email{paddy@iucaa.ernet.in}
\affiliation{IUCAA,
Post Bag 4, Ganeshkhind, Pune - 411 007, India\\}

\date{\today}


\begin{abstract}

It was shown recently that, in the case of Schwarschild black hole, one can obtain the correct
thermodynamic relations by studying a model quantum system and using a particular duality transformation. We study this approach further for the case a general spherically symmetric horizon. We show that the idea works for a  general case only if we define the entropy $S$ as a congruence (``observer") dependent quantity and the energy $E$ as the integral over the source of the gravitational acceleration for the congruence. In fact, in this case, one recovers the relation $S=E/2T$ between entropy, energy and temperature previously proposed by one of us in gr-qc/0308070. This approach also enables us to calculate the quantum corrections of the Bekenstein-Hawking entropy formula for all spherically symmetric horizons.
\end{abstract}

\maketitle
\vskip 0.5 in
\noindent
\maketitle

There is an intriguing analogy between gravitational dynamics of horizons  and thermodynamics
(for a recent review, see ref. \cite{paddyR}).
We do not have fundamental understanding of this though most people believe this indicates a deep and as yet undiscovered aspect of quantum gravity. The first results, of course, were that of black hole horizon \cite{hawkings}, Rindler horizon \cite{daviesunruh} and De-Sitter horizon \cite{gh}. In the case of these (and other horizons) one can associate a temperature fairly unambiguously. In the case of blackhole horizon, one also associates an entropy. In the case of Rindler and De-Sitter horizon, the observer dependence of the horizon makes people uneasy as regard associating entropy to the horizon (and  most people try not to take a clear stand on this matter!).
On the other hand there are strong arguments suggesting \textit{all} horizons have thermodynamic variables associated with them and  entropy of horizons \textit{is} an observer dependent construct \cite{entropyobs}.
If so, one can associate thermodynamic laws with all horizons and -- in fact -- this appears to offer
an entirely new perspective on gravity
  \cite{paddy,holo}.

In conventional  systems, one can derive the laws of thermodynamics from a more fundamental theory --- statistical mechanics. In such systems, entropy  can be defined as the logarithm of total number of accessible microstates corresponding to the same macrostate. The existence of an entropy associated with any horizon provides a strong motivation for one to look for certain microstates of the underlying quantum theory.  Although there have been several attempts to derive black hole entropy formula from counting the possible microstates in both string and loop formalisms, a comprehensive understanding of this issue remain elusive.

Given this state of affairs, there is justification to  study various aspects of the horizon thermodynamics and try to construct phenomenological models based on them.
Recently, Balazs et. al.\cite{balaz} has proposed an intriguing model in which they have considered a ``dual thermodynamics'' corresponding to isolated Schwarzschild black hole
 and have tried to obtain the entropy from a dual theory, to which standard statistical mechanics is applicable. They show that the standard (black hole) thermodynamic relations are invariant under the
 transformations $E' \to A/4$, $S' \to M$, and $T' \to T^{-1}$ where $A$, $M$ and $T$ are the horizon area, mass and Hawking temperature of a black hole and $E'$, $S'$ and $T'$ are the energy, entropy and temperature of the corresponding dual quantum system \cite{balaz}. After working out the standard thermodynamics of the dual system, they apply the inverse transformations to get standard horizon thermodynamics as well as the logarithmic corrections to the original Bekenstein-Hawking entropy formula. This approach seem to provide a description of a strongly interacting gravitational system like black hole in terms of a weakly interacting quantum mechanical dual system. 
 
 The result is sufficiently intriguing that one would like to understand its origin and domain of applicability. In particular one would like to know whether it generalizes for an arbitrary horizon (like e.g., De-Sitter horizon, which will play an important role in the study of cosmological constant \cite{cc}). We will address several aspects of this question in this paper in an attempt to understand this result.

 Any attempt to generalize these result beyond the Schwarzschild blackhole raises an operational issue: While one can define the temperature for a sufficiently general class of horizons (Schwarschild, Reissner-Nordstrom, De-Sitter, Rindler .....), and even take entropy per unit transverse area to be $(1/4)$ (so that  non-compact horizons are also taken care of) it is not easy to define the \textit{energy} associated with arbitrary horizons! Even for  Reissner-Nordstrom blackhole there are different expressions for energy available in the literature (see e.g., \cite{Mustafa,Elias} and references cited therein).
 Fortunately, for all spherically symmetric metrics, there is a natural way of defining this quantity. This is explained in detail in ref.\cite{paddy} and is essentially $|U|=a/2$ in geometric units where $a$ is the horizon radius. However, we shall show below that,  with this definition, one can \textit{not} generalize the idea of duality to spherically symmetric spacetime in a consistent manner. 
 
 On the other hand, one can define another expression for energy $E$ which is the source of gravitational acceleration using the Tolman-Komar integral (see \cite{paddy2,tolman}). It was shown in ref.\cite{paddy2} that this expression for energy also arises naturally in the case of any static horizon and allows one to obtain an ``equation of state" between entropy, temperature and energy in the form $S=E/2T$. Surprisingly, this result arises directly from the duality model! We shall now describe these results in detail and provide a brief discussion.

 Consider a static, spherically symmetric horizon, in the spacetime  described by a metric (in which we have adopted natural units, with $\hbar = c = G = 1$):
\begin{equation}
ds^2 = f(r) dt^2 - \frac{1}{f(r)} dr^2 -r^2 d\Omega^2.
\end{equation}
We assume that the function $f(r)$ has a simple zero at $r=a$ with finite $f'(a)$ so that the spacetime has a horizon at $r=a$. Periodicity in Eulerian time  allows us to associate a temperature $ T = f'(a)/ 4 \pi $ with the horizon. (Even for spacetimes with multi-horizons this prescription is locally valid for each horizon surface \cite{trith,paddy}). The Bekenstein-Hawking entropy is equal to the one quarter of the horizon area  and therefore is given by,
\begin{equation}
S = \frac{1}{4} ~(4 \pi a^2) = \pi a^2. \label{entropy}
\end{equation} 
These correspondences have been worked out in detail in \cite{paddy}; when $f'(a)<0$, like in the case of De-Sitter, one needs to make appropriate sign changes, which is also explained in \cite{paddy}. For our purpose we will ignore these complications.

The real ambiguity is in giving a proper prescription for the energy --- a well known problem in general relativity. For the Schwarzschild black hole it is obviously equal to the total mass $M$, but for  De-Sitter spacetime or even in the apparently simpler (asymptotically flat) case of
a charged black hole, the definition of energy is non-trivial. We will handle this difficulty by not assuming a-priori  any  prescription for energy $E$; instead we will find its generic form if the duality principle is valid. Now, the duality transformation considered by in ref.\cite{balaz} is ,
\begin{eqnarray}
S \to  E,~~~E \to S,~~~ T \to \frac{1}{T}~~~\textrm{and}~~~\frac{\mu}{T} \to -\mu.
\end{eqnarray}
Where $\mu$ is the chemical potential. Such a transformation will preserve the form of the first law of thermodynamics for both the black hole and the dual system. If we  assume its validity and map a general spherically symmetric horizon to its dual system, we get the expressions for energy, entropy and temperature  for the dual system to be:
\begin{eqnarray}
S_{d} = E, ~~~E_{d} = \pi a^2,~~~\textrm{and}~~~ T_{d} = \frac{4 \pi}{f'(a)}.
\end{eqnarray}
(Note that we are not assuming any specific form for the energy $E$ of the horizon.) Taking the dual system as a one dimensional Bose gas, the expressions for energy and entropy are given by \cite{balaz},
\begin{eqnarray}
E_{d} = p L &=&\frac{1}{12} g L \pi  T_{d}^2
   - \frac{ g L}{2 \pi } \left(\log \left(-\mu_{d} /T_{d} \right)-1\right) T_d \mu_d
   - \frac{g L}{8\pi } \mu_d  ^2
   + O\left(\mu_d ^3\right),
    \\
     S_d &=&\frac{1}{6} g L \pi T_d
   - \frac{g L}{2 \pi } (\log (-\mu_{d}/T_d ) - 1) \mu_d
   + O\left(\mu_d ^3\right).
\end{eqnarray}
(One can easily perform all the calculations for the Fermi gas.) The coefficient $g$ is the internal degrees of freedom of the gas particles and $L$ is the size of the system. Consistency requires that the leading order term of the entropy of the dual system gives correct value of of the black hole energy under the reverse transformation. This fixes the value of the quantity $g L$ is uniquely as $g L = 3 E f'(a)/2 \pi^2$,
With this value of $gL$, the \textit{leading term} in the expression of energy $E_d$ of the dual system becomes,
\begin{equation}
E_{d} = \frac{2 \pi E  }{f'(a)}.
\end{equation}
If under a reverse transformation this expression gives the correct black hole entropy, one must have, $\pi a^2 = 2 E \pi /f'(a)$ thereby allowing us to determine the energy associated with the horizon
\begin{equation}
 E = \frac{a^{2} f'(a)}{2}. \label{energy}
\end{equation}
The above analysis is exactly the same as the one performed in ref.\cite{balaz} but for general context, keeping $E$ unspecified but $S=\pi a^2,T = f'(a)/ 4 \pi$. We stress that we have not done anything drastic or unconventional except to adhere to the duality prescription.

For Schwarzschild spacetime $a = 2M$, and $f'(a) = 1/2M$ and the above equation will give correct value of energy $E = M$ if we have $a f'(a) = 1$. This condition is trivially satisfied for Schwarzschild blackhole. This is essentially the result of \cite{balaz}.

Let us next try Reissner-Nordstrom blackhole. Then $f(r) = \left(1 - 2 M/r + Q^2/r^2  \right)$ and
the horizon is  at $a = M + \sqrt{M^2 - Q^2}$.
The temperature associated with this outer horizon  is $ T =\kappa /4 \pi $ where $\kappa$ is the surface gravity of the outer horizon. In this case it is easy to see that $a f'(a) = 2 M/a - 2 Q^{2}/a^{2} $
(so that $a f'(a)  \neq 1$ unless $Q=0$). If we require this analysis to be applicable for charged black holes also, the energy $E$ should be given by Eq.~(\ref{energy}):
\begin{eqnarray}
E = \frac{a^{2} f'(a)}{2} = M - \frac{Q^{2}}{a} = \sqrt{M^2 - Q^2}. \label{chargedenergy}
\end{eqnarray}
In the literature several expressions are given for energy of Reissner-Nordstrom metric and this one corresponds to Moller energy  for  Reissner-Nordstrom black hole \cite{Mustafa,Elias}.
However, since our motivation is to study the thermodynamics of the horizons, any definition of energy should come from a thermodynamic consideration. But the expression of the energy in Eq.~(\ref{chargedenergy}) does not agree with this prescription. For example, if we drop an uncharged particle of mass $M$ into a Reissner-Nordstrom blackhole, its energy should change by $dM$ and the resulting change in entropy should be related to this by $dE=TdS$. But if we start with the expression of energy in Eq.~(\ref{chargedenergy}), then the change in mass will result in a change in energy $dE = M dM/ \sqrt{M^2 - Q^2}$. On the other hand $TdS= \left(M/a - Q^2/a^{2}\right)da = dM$ which does not match $dE$. Therefore the energy expression in Eq.~(\ref{chargedenergy}) \textit{does not have a proper thermodynamic interpretation}.

There is, however, another interesting interpretation possible for this result, which is based on the formalism developed in ref.\cite{paddy2}. We shall briefly recall this result, which is
based on the point of view that the entropy associated with any horizon is due to the information which are hidden by the horizon.
In order to formulate this idea mathematically, we need to set up the geometrical framework which is adapted to a congruence of observers who sees a horizon. The metric of a static spacetime can always be put in the form $ds^2 = - N^2 dt^2 + \gamma_{\mu \nu} dx^{\mu} dx^{\nu}$, where $N$ and $\gamma_{\mu \nu}$ are independent of time $t$ (Greek indices cover 1,2,3 and Latin indices cover 0-3). The comoving observer at $x^{\mu} = $constant have the four velocity $u_{i} = - N \delta^{0}_{i}$ and the four acceleration $ a^{i} = (0, \partial ^{\mu}N/N)$. If $N \to 0$ on a two-surface and $N a = (\gamma_{\mu \nu} \partial ^{\mu}N \partial ^{\nu}N )^{1/2}$ is finite (say $\kappa$, the surface gravity), then the coordinate system has a horizon.  Regularity in the Euclidean sector requires the periodicity in Euclidean time with the period $ | \beta | = 2 \pi / \kappa $, allowing us to define a temperature $T = | \beta ^{-1} |$ in terms of the derivatives of $N$, whenever there is a horizon. The expression for entropy associated with this horizon is given as \cite{paddy2},
\begin{eqnarray}
S = \frac{1}{8 \pi G} \int \sqrt{-g} ~d^{4}x \nabla_{i} a^{i}. \label{def_entropy}
\end{eqnarray}
Obviously, the entropy defined by the above expression depends on the choice of the congruence, through the four vector $u^{i}$, but is \textit{generally covariant}. (For justification behind  this definition of gravitational entropy, see ref.\cite{paddy2}). Now, in any spacetime, there is a differential geometric identity \cite{mtw},
\begin{eqnarray}
R_{bd} u^{b} u^{d} = \nabla _{i} (K u^{i} + a^{i} ) - K_{ab} K^{ab} + K_{a}^{a} K_{b}^{b}.
\end{eqnarray}
where $K_{ab}$ is the extrinsic curvature of spatial hypersurfaces and $K$ is its trace. In static spacetime we have $K_{ab} = 0$ and when combined with Einstein's equation we can write:
\begin{eqnarray}
\frac{1}{8 \pi G} \nabla_{i} a^{i} = ( T_{ab} - \frac{1}{2} T g_{ab}) u^{a} u^{b}. \label{source}
\end{eqnarray}
This equation relates the integrand of Eq.~(\ref{def_entropy}) to the matter stress tensor $T_{ab}$.
We next note that the source for gravitational acceleration is the covariant combination $ (T_{ab}- \frac{1}{2} T g_{ab}) u^{a} u^{b}$, and the corresponding energy $E$ is given by Tolman-Komar integral \cite{tolman},
\begin{eqnarray}
E = 2 \int_{{\mathcal V}} d^{3}x \sqrt{\gamma} N (T_{ab}- \frac{1}{2} T g_{ab}) u^{a} u^{b}. \label{Energy}
\end{eqnarray}
From Eq.~(\ref{def_entropy}), Eq.~(\ref{energy}) and Eq.~(\ref{source}), we can easily find that,
\begin{equation}
S = \frac{1}{2} \beta E . \label{relation}
\end{equation}
A closer inspection of the Eq.~(\ref{energy}) and Eq.~(\ref{entropy}) reveals that there exits a exactly the same  relationship between entropy $S$ and energy $E$ defined by the dual system approach!
 Therefore, the natural conclusion of our analysis is that, the generalization of this duality transformations in \cite{balaz} for any spherically symmetric horizon is \textit{only possible} when the entropy and the energy associated with the horizon are defined in accordance with ref.\cite{paddy2}. That is, the energy-entropy duality in general should be taken as a dual transformation between the \textit{entropy $S$ in Eq.~(\ref{def_entropy}) and energy $E$ in Eq.~(\ref{Energy})}, so that the relationship  Eq.~(\ref{relation}) will always be satisfied. This will ensure the consistency of the entire approach and give the correct value of the Bekenstein-Hawking entropy under reverse transformation.

For the sake of completeness we also calculate quantum corrections to the horizon entropy
for a general spherically symmetric horizon, using
corrections to the energy of the dual quantum gas. The energy of the  Bose model is
\begin{eqnarray}
 E_d = p L &=&\frac{1}{12} g L \pi  T_{d}^2
   - \frac{ g L}{2 \pi } \left(\log \left(-\mu_{d}/T_{d} \right)-1 \right) T_d \mu_{d}
   - \frac{g L}{8\pi } \mu_{d} ^2
   + O\left(\mu_{d} ^3\right).
\end{eqnarray}
Using $g L=3 E f'(a)/2 \pi^2$, and performing the reverse transformation, the entropy of the horizon is obtained as,
\begin{eqnarray}
S = \pi a^2 + \frac{6 a^2}{\pi} \left(\log{\mu} - 1 \right) \mu - \frac{3 a^2}{2 \pi} \mu^{2} +O\left(\mu ^3\right).
\end{eqnarray}
Quantum corrections to the entropy of various black holes were determined using the Cardy formula \cite{Cardy,Das}. In order to obtain the correct coefficient of the logarithmic quantum correction term \cite{Kaulwe} we need to fix the value of $\mu$ as,
\begin{eqnarray}
\mu = \frac{\pi}{4 a^2}
\end{eqnarray}
With this choice of $\mu$ the entropy of the horizon becomes,
\begin{eqnarray}
S =\pi a^2 - \frac{3}{2} \log{(\pi a^2)} - \frac{3}{2} \left( 1+ \log \left( \frac{4}{\pi^2}\right) \right) + ....
\end{eqnarray}
For Schwarzschild case $a = 2 M$, and we recover the correct expression for the black hole entropy with logarithmic quantum corrections as shown in \cite{balaz}. 
The actual value is, of course, obtained by adjusting the free parameter $\mu$ in the theory and hence is not so important; but the fact that the nature of the corrections have the logarithmic form is interesting and is worthy of futher investigation. 
If the duality ideas are correct, these terms are also universal and is applicable to all horizons.

Our analysis clearly shows that it is indeed possible to generalize the duality principle in \cite{balaz} for a generic spherically symmetric horizon provided the definition of entropy and energy are in accordance with the results obtained in ref. \cite{paddy2}. This suggests that the approach of understanding the horizon thermodynamics based of a duality transformation is possibly quite generic and therefore may be important for the understanding of the underlying quantum theory.

One of the author (S.S.) is supported by the Council of Scientific \& Industrial Research, India.

\end{document}